\documentclass[doublecol]{epl2}
\title{Doping effect on thermoelectric properties of MoS$_2$ }
\shorttitle{Title} %Insert here a short version of the title if it exceeds 70 characters

\author{Huaihong Guo\inst{1} \and Teng Yang\inst{1}\thanks{E-mail: \email{yangteng@imr.ac.cn}} \and Peng Tao\inst{1} \and Zhidong Zhang\inst{1}}
\institute{
\inst{1} Shenyang National Laboratory for Materials Science,
Institute of Metal Research and International Centre for Materials Physics,
Chinese Academy of Sciences, 72 Wenhua Road, Shenyang 110016,
PRC
}

\abstract{We systematically study thermoelectric properties of layered MoS$_2$ by doping, based on Boltzmann transport theory and first-principles calculations. We obtain optimal doping region (around 10$^{19}$ cm$^{-3}$) by looking closely to the temperature and doping level dependent thermopower, electrical conductivity, power factor (PF) and ultimately figure of merit (ZT) coefficient along in-plane and cross-plane directions. MoS$_2$ has a vanishingly small anisotropy of thermopower but a big anisotropy of electrical conductivity and electronic thermal conductivity in optimal doping region. $\kappa_e$ is comparable to $\kappa_l$ in the plane while $\kappa_l$ dominates over $\kappa_e$ across the plane. ZT can reach as high as 0.3 at around 700 K. In-plane direction is demonstrated to be more preferable for thermoelectric applications of MoS$_2$ by doping.}

\begin{document}

\maketitle

\section{Introduction}
Thermoelectrics play a key role for power generation and refrigeration \cite{application1, application2, application3}. Chalcogenide composite materials, such as Bi$_{2}$Te$_{3}$, PbTe and others \cite{Tritt,highTTE1,PRB.83.115110,hightTTE3,Bi2Te3andSb2Te31,Bi2Te3andSb2Te32,Bi2Te3andSb2Te33}, have attracted much research interest for decades because of their high performance on thermoelectric conversion. Transition-metal dichalcogenide MoS$_2$, one prototype material in chalcogenide family, has distinctive electronic, mechanical, catalytic and tribological properties \cite{tenne,PhysRevB.48.10583,catalyst,photocatalysis,electronic1, nnature.6.147,FET,solarcell,hstorage, battery}. Recent research on its optical properties \cite{PhysRevLett.105.136805} and lattice dynamics \cite{ACSNano.4.2695,lattdyn} has aroused renewed interest but its thermoelectric properties have barely been studied \cite{jap-hhguo}. Nevertheless, the lowest thermal conductivity obtained experimentally in a MoS$_2$-related structure \cite{science315.351}, together with an unusually large thermopower \cite{JPCS1983, SSC1986, PPSB1953} found in MoS$_2$, may render it a promising candidate for thermoelectric application.

Unfortunately, available experimental data concerning the thermoelectric transport properties of MoS$_2$ are quite scarce and fragmentary in literature. The most significant experimental work was performed by Mansfield and Salam \cite{PPSB1953} and Thakurta et. al. \cite{JPCS1983}, who studied temperature dependence of the electrical transport properties including thermopower and electrical conductivity at only three samples with low dopings, but failed to investigate on the thermal-related properties. Kim et. al. \cite{kappa2} merely worked on the thermal conductivity $\kappa$. Based on these incomplete experimental data, it is difficult to estimate the figure of merit ZT coefficient which is crucial to evaluate thermoelectric conversion capability. Moreover, available extrinsic carrier concentrations in MoS$_2$ single crystals were quite low. It remains a big challenge to obtain a wide doping region in experiment but nonetheless is essential to tuning carrier density appropriate for an optimum ZT.

In this letter, we study theoretically the thermoelectric transport properties of MoS$_2$ over a range of doping level (from 10$^{15}$ to 10$^{20}$ cm$^{-3}$) for an optimization of its thermoelectric conversion efficiency. We have the following findings: (1) Thermopower is attainable more than 200 $\mu$V/K over a wide range of dopings and directional anisotropy vanishes between in-plane and cross-plane thermopowers for doping level above 10$^{17}$ cm$^{-3}$. (2) Anisotropic electronic scattering time exists between in-plane and cross-plane directions, which accounts for two orders of magnitude difference between electrical conductivities $\sigma_{xx}$ and $\sigma_{zz}$ and also gives rise to anisotropy between electronic thermal conductivity $\kappa^{xx}_e$ and $\kappa^{zz}_e$. (3) In optimal doping region, $\kappa_e$ is comparable to $\kappa_l$ in the plane while $\kappa_l$ dominates over $\kappa_e$ across the plane. (4) Carrier density around 10$^{19}$ cm$^{-3}$ is sufficient for optimal thermoelectric performance, figure of merit ZT coefficient reaches 0.3 at 700K within the plane. (5) A preference for in-plane over cross-plane direction by doping for thermoelectric applications of MoS$_2$ is demonstrated.

%---------------------------------------------------------------------
\begin{figure}[t]
\includegraphics[width=1.0\columnwidth]{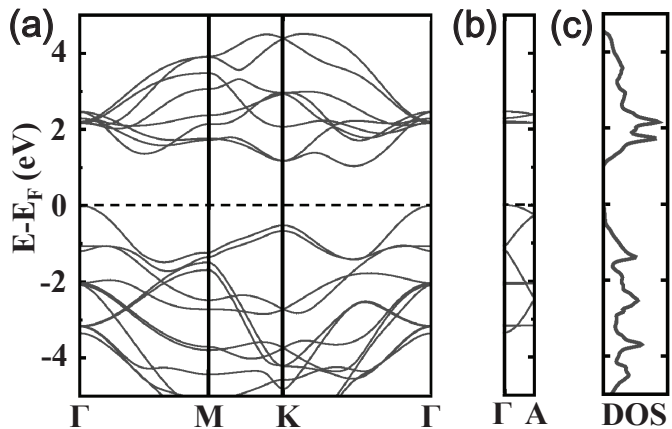}
\caption{(Color online) EV-GGA electronic band structure of MoS$_2$ along high-symmetry lines (a) $\Gamma$-M-K-$\Gamma$ in plane and (b) $\Gamma$-A across plane in the hexagonal Brillouin zone (BZ). The valence-band edge is set as zero and marked with a dashed line. Electronic density of states in (c) shows a comparatively higher value at the conduction band edge than at the valence band edge.
\label{Fig1} }
\end{figure}
%---------------------------------------------------------------------

\section{METHODOLOGY}
The band structure of MoS$_2$ is calculated by using the general potential linearized augmented plane-wave(LAPW) method as implemented in the WIEN2K package \cite{wien2k}. The electronic exchange-correlation is described within the generalized gradient approximation (GGA) of Perdew-Burke-Ernzerhof(PBE) flavor \cite{PRL.77.3865}. We use 5000 k points in the full Brillouin zone(BZ) to achieve a total energy convergence better than 1 meV/atom. MoS$_2$ has P$_{63}$/mmc space group symmetry and consists of a hexagonal plane of molybdenum atoms sandwiched by two hexagonal planes of sulfur atoms. The unit cell contains two alternating and weakly van-der-Waals-bonded layers with an AB stacking along c axis. The experimental lattice parameters \cite{explatt} (a = 3.16$\AA$, c = 12.295$\AA$) are used here.

We calculate transport properties based on Boltzmann transport theory applied to the band structure. In the following part, we discuss the dependence of transport functions including thermopower S, electrical conductivity $\sigma$, power factor(PF), and ultimately figure of merit coefficient ZT on temperature and doping level along two perpendicular directions. The electron scattering time is assumed to be independent of energy due to its good description of S(T) in a number of thermoelectric materials \cite{PRB.83.115110, PRB.83.245111, PRB.80.075117}. In this sense, no adjustable parameters are needed to calculate those transport functions. The integration is done within the BOLTZTRAP transport code \cite{PRB.68.125212}. A very dense mesh with up to 18000 k points in the BZ is used.
%---------------------------------------------------------------------
\begin{figure}[t]
\includegraphics[width=1.0\columnwidth]{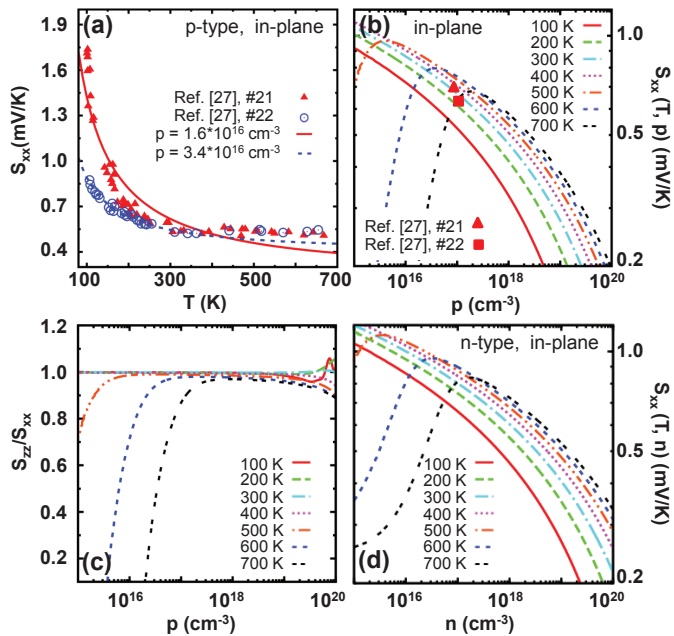}
\caption{(Color online) (a) Temperature dependence of calculated in-plane thermopower S$_{xx}$ of MoS$_2$, compared with experimental data by Mansfield and Salam \cite{PPSB1953} at two doping levels p = 1.6 $\times$ 10$^{16}$ (filled triangle) and 3.4 $\times$ 10$^{16}$ (empty circle) holes per cm$^{3}$. Doping level dependence of (b) in-plane thermopower S$_{xx}$(p, T), (c) ratio of cross-plane S$_{zz}$(p, T) over in-plane S$_{xx}$(p, T) and (d) in-plane S$_{xx}(n, T)$ at different temperatures. The temperature ranges from 100K to 700K for some practical reason. Hole and electron doping are respectively used in (a-c) and (d). Experimental data with p = 7.6 $\times$ 10$^{16}$ and 1.0 $\times$ 10$^{17}$ cm$^{-3}$ at 200K by Mansfield and Salam \cite{PPSB1953} is respectively marked by filled triangle and square in (b).
\label{Fig2} }
\end{figure}
%---------------------------------------------------------------------

\section{RESULTS AND DISCUSSION}
An improved band gap is essential for calculating transport properties, so the Engel-Vosko GGA (EV-GGA) formalism \cite{PRB.47.13164} is applied to calculate the band gap more accurately. In Fig.~\ref{Fig1}, we present our calculated band structure and density of states of MoS$_2$. The in-plane and cross-plane cases are considered separately in Fig.~\ref{Fig1}(a, b). In Fig.~\ref{Fig1}(a), an indirect gap of 1.04 eV is obtained between top of valence band at $\Gamma$ and bottom of conduction band at one k point from K to $\Gamma$. A similar band structure calculated from standard PBE-GGA formalism gives an indirect gap of 0.84 eV, which agrees with reported value \cite{prb.84.045409}. Compared with $\Delta_i$ $\sim$ 1.20 eV from experiment \cite{expgap2}, it is clear that EV-GGA does improve the band gap calculation upon PBE-GGA. In contrast to the pronounced dispersive in-plane electronic bands shown in Fig.~\ref{Fig1}(a), the cross-plane bands shown in Fig.~\ref{Fig1}(b) are quite flat, showing a very weak cross-plane bonding due to Van der Waals interaction. The structural anisotropy induces an anisotropy between in-plane and cross-plane band gaps. The calculated cross-plane gap is found to be 2.20 eV.

Besides the band gap anisotropy, a strong asymmetric feature of band structure between valence and conduction bands implies that the thermoelectric properties of n-type MoS$_2$ would be very different from that of p-type. The heavy and doubly degenerate bands near the conduction-band minimum suggest that the n-type MoS$_2$ would have better thermoelectric performance. In Fig.~\ref{Fig1}(c), total density of states (DOS) also shows this preference, there is a comparatively higher DOS very close to the conduction band edge than that near the valence band edge. Considering more experimental data are found for p-type in literature \cite{SSC1986,PPSB1953,JPCS1983}, we focus on hole-doped MoS$_2$ in this study.
%---------------------------------------------------------------------
\begin{figure}[t]
\includegraphics[width=1.0\columnwidth]{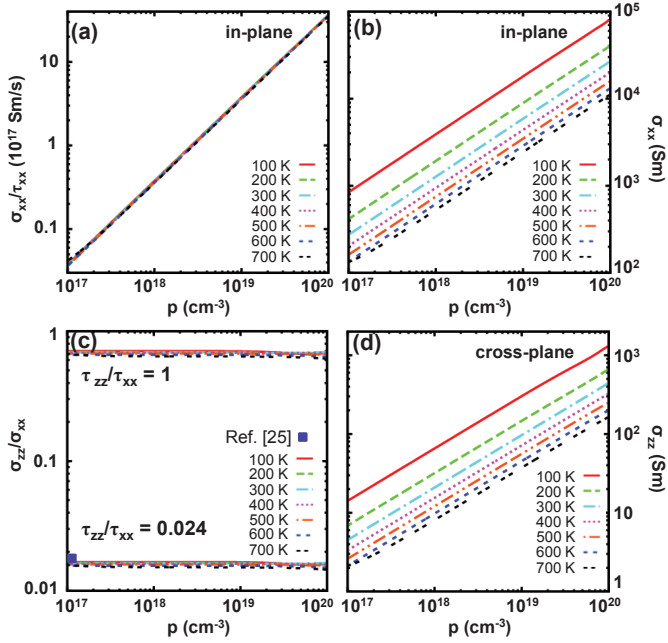}
\caption{(Color online) Doping level dependence of (a) in-plane $\sigma_{xx}/\tau_{xx}$, (b) in-plane electrical conductivity $\sigma_{xx}$, (c) ratio of cross-plane $\sigma_{zz}/\tau_{zz}$ over in-plane $\sigma_{xx}/\tau_{xx}$ and (d) cross-plane $\sigma_{zz}$, at different temperatures. Isotropic and anisotropic electronic scattering time $\tau$ are respectively assumed in (c), for instance, $\tau_{zz}/\tau_{xx} = 1$ and 0.024 with the anisotropic one fitting well the experimental data in blue square from Thakurta et. al. \cite{JPCS1983} and thereby used to derive $\sigma_{zz}$ in (d).
\label{Fig3} }
\end{figure}
%---------------------------------------------------------------------

In-plane and cross-plane thermopower are firstly studied. We initially compare our calculated S with experimental data from Mansfield and Salam \cite{PPSB1953} in Fig.~\ref{Fig2}(a,b). Samples in their experiment were p-type from nature with hole carrier concentration as low as 10$^{15}$ $\sim$ 10$^{17}$ cm$^{-3}$. A very good agreement is obtained. Both our calculation and available experimental data show a value of in-plane thermopower S higher than 400 $\mu$V/K and S decreases with increasing temperature as shown in Fig.~\ref{Fig2}(a). Then we extend our discussion to high doping of 10$^{17}$ $\sim$ 10$^{20}$ cm$^{-3}$ (corresponding to p = 10$^{-5}$ $\sim$ 10$^{-2}$ holes per unit cell in our case) where thermoelectric properties are expected to be optimized as predicted from theory \cite{mahan1998} and experimentally observed in many materials \cite{JMSL.14.617,PRL.106.206601,nmaterial}. It also applies in MoS$_2$ as we will show later. From Fig.~\ref{Fig2}(b), the thermopower in the high doping region, though decreasing with doping level, takes a value of at least 200 $\mu$V/K (ZT$>$2.4 is expected from the Wiedemann-Franz law.) and increases with increasing temperature. To see a possible anisotropy between in-plane and cross-plane S, we then show the ratio of S$_{zz}$ over S$_{xx}$ in Fig.~\ref{Fig2}(c). A relatively high anisotropy of thermopower below 10$^{17}$ cm$^{-3}$, possibly due to difference of effective mass (or mobility) and density between band-edge carriers moving along two directions, according to the Mott formula\cite{jziman} S $\sim$ $\frac{\partial \ln (n \mu)}{\partial E}$ where n, $\mu$ are respectively carrier density and mobility. Anisotropy vanishes as doping goes above 10$^{17}$ cm$^{-3}$ and both S$_{xx}$ and S$_{zz}$ show similar magnitude and dependence on hole doping level and temperature, which is different from layered conductive thermoelectric oxides \cite{PRL.104.176601,APL.98.202109}. Finally, to confirm that n-type MoS$_2$ may be better, we briefly compare n-type in Fig.~\ref{Fig2}(d) with p-type in Fig.~\ref{Fig2}(b). Expectedly, bigger value of thermopower of n-type MoS$_2$ than of p-type is found.

%---------------------------------------------------------------------
\begin{figure}[t]
\includegraphics[width=1.0\columnwidth]{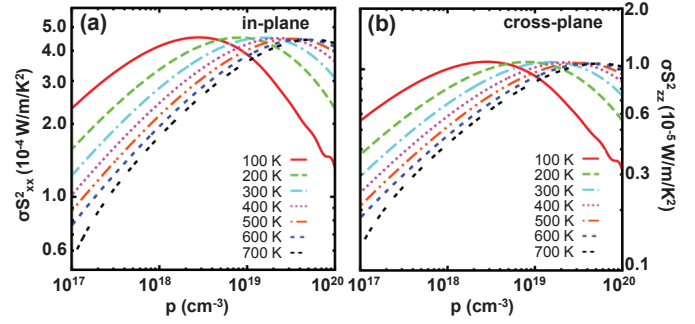}
\caption{(Color online) Doping level dependence of (a) power factor $\sigma_{xx} S^2_{xx}$ and (b) $\sigma_{zz} S^2_{zz}$ of p-type MoS$_2$ at different temperatures.
\label{Fig4} }
\end{figure}
%---------------------------------------------------------------------
Based on the energy-independent scattering time approximation, it is quite straightforward to get doping dependence of $\sigma$/$\tau$ at various temperature from the electronic band structure, upon which we calculate $\sigma$ if $\tau$ is known. In Fig.~\ref{Fig3}(a), we show calculated $\sigma_{xx}$/$\tau_{xx}$ depending on hole doping and temperature. We find an almost temperature independence but an approximately linear doping level dependence of $\sigma_{xx}$/$\tau_{xx}$, namely $\sigma_{xx}$/$\tau_{xx}$ $\sim$ T$^0$ p. For a quadratic band dispersion in an electron-phonon approximation, $\sigma$ $\sim$ p$^{\frac{2}{3}}$ T$^{-1}$ stands, this results in $\tau_{xx}$ $\sim$ T$^{-1}$p$^{-\frac{1}{3}}$, in consistent with the analytical treatment of carriers scattered by lattice vibrations in a semiconductor \cite{jziman,PRB.83.115110}. To calculate $\sigma_{xx}$, we need some experimental input to get $\tau_{xx}(T, p)$. Here we use $\sigma_{xx}$ = 0.16 $\Omega^{-1} cm^{-1}$ at 100 K and 1.4$\times$10$^{15}$ cm$^{-3}$ doping level from the experimental data by Thakurta et. al. \cite{JPCS1983} and get $\tau_{xx}$ = 3.04 $\times$ 10$^{-6}$ T$^{-1}$p$^{-\frac{1}{3}}$ with Kelvin and cm$^{-3}$ as the unit of T and p, respectively. Plugging it into our calculated $\sigma$/$\tau$, we show $\sigma_{xx}(T, p)$ in Fig.~\ref{Fig3}(b), and power factor $\sigma_{xx} S^2_{xx}(T, p)$ in Fig.~\ref{Fig4}(a).

Electrical conductivity along c axis is also calculated and a strong anisotropy is found between the in-plane and cross-plane carrier scattering time $\tau$. When isotropic $\tau$ is assumed, namely, $\tau_{zz}/\tau_{xx}$ = 1, we obtain $\sigma_{zz}$/$\sigma_{xx}$ close to unity in Fig.~\ref{Fig3}(c), which is against the reported result \cite{JPCS1983}. To fit $\sigma_{xx}$/$\sigma_{zz}$ of two orders of magnitude in experiment, we use anisotropic $\tau_{zz}/\tau_{xx}$ = 0.024, which suggests that a strong anisotropy of carrier scattering time should play a role in this system, and therefore obtain $\tau_{zz}$ = 7.30 $\times$ 10$^{-8}$ T$^{-1}$p$^{-\frac{1}{3}}$. Finally, we are able to calculate $\sigma_{zz}(T, p)$ and show it in Fig.~\ref{Fig3}(d).

With thermopower and electrical conductivity available, we are able to evaluate power factor. For an optimized thermoelectric performance, peak value of PF, the corresponding doping level and temperature are more concerned here. Both power factors along two perpendicular directions have peak values spanning in a wide doping range from 10$^{17}$ to 10$^{20}$ cm$^{-3}$. From Fig.~\ref{Fig4}, the value of peak power factor is nearly constant, while its temperature increases with the hole doping level. Due to the anisotropic carrier scattering time $\tau$, nearly 50-fold difference is found between in-plane and cross-plane power factors, e.g., $\sigma$$S^2_{xx}$ and $\sigma$$S^2_{zz}$ maxima at 700 K are respectively 4.1$\times$10$^{-4}$ W/m/K$^2$ and 1.0$\times$10$^{-5}$ W/m/K$^2$. The in-plane PF value is close to that of good thermoelectric materials \cite{mahan1998}.
%---------------------------------------------------------------------
\begin{figure}[t]
\includegraphics[width=1.0\columnwidth]{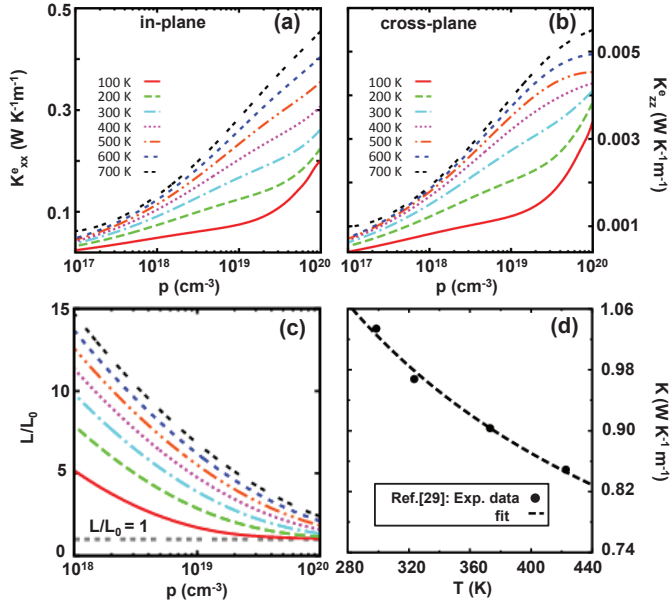}
\caption{(Color online) (a) In-plane electronic thermal conductivity $\kappa^e_{xx}$, (b) cross-plane $\kappa^e_{zz}$ and (c) in-plane L/L$_0$ as function of doping and temperature. L = $\kappa$/($\sigma$ T) and L$_0$ is the Lorenz number 2.44$\times$10$^{-8}$ W$\Omega$K$^{-2}$. (d) Temperature dependence of thermal conductivity $\kappa$ from experiment by Kim et. al. \cite{kappa2}, which are presented by filled circles and fitted by $\kappa$ = 183.103/T + 0.412671 in dashed line.
\label{Fig5} }
\end{figure}
%---------------------------------------------------------------------

%---------------------------------------------------------------------
\begin{figure}[t]
\includegraphics[width=1.0\columnwidth]{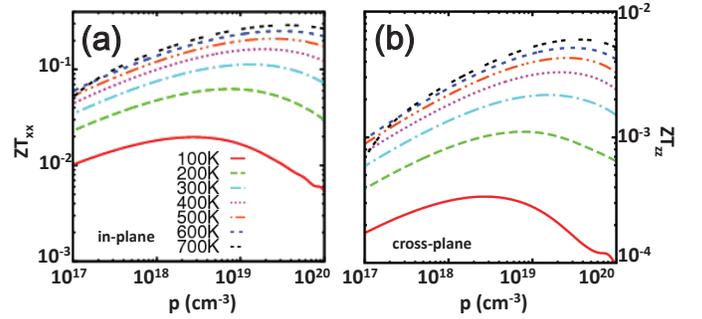}
\caption{(Color online) Doping level dependent (a) in-plane and (b) cross-plane figure of merit coefficient ZT of p-type MoS$_2$ at different temperatures. The experimental thermal conductivities $\kappa$ from Fig.\ref{Fig5}(d) were used.
\label{Fig6} }
\end{figure}
%---------------------------------------------------------------------

To optimize ZT value, it is also essential to know thermal conductivity $\kappa$, including electronic $\kappa_e$ and lattice $\kappa_l$. Based on the scattering time approximation previously discussed, we firstly calculate electronic thermal conductivity and show it in Fig.~\ref{Fig5}(a,b). $\kappa_e$ increases with carrier density p and temperature. At least two orders of magnitude difference between $\kappa^e_{xx}$ and $\kappa^e_{zz}$ are found, which is consistent with electrical conductivity case. Usually one gets $\kappa_e$ from electrical conductivity $\sigma$ by using the Wiedemann-Franz law, namely $\kappa_e$/($\sigma$ T) = 2.44$\times$10$^{-8}$ W$\Omega$K$^{-2}$, the so-called  Lorenz number. However, it seems not the case here. In Fig.\ref{Fig5}(c) we normalize L(=$\kappa_e$/($\sigma$ T)) by the Lorenz number and plot it as function of doping and temperature. L/L$_0$ is close to one as carrier density goes beyond 10$^{20}$ cm$^{-3}$. Unlike the electronic $\kappa_e$, the lattice $\kappa_l$ can't be calculated from electronic band structure. Here we simply use experimental data from literature \cite{kappa2} and show it in Fig.~\ref{Fig5}(d). Kim et. al. \cite{kappa2} measured temperature dependence of $\kappa$ for MoS$_2$ sample as shown by filled circles in Fig.\ref{Fig5}(d). The lattice thermal conductivity dominates in the cross-plane direction, with two orders of magnitude bigger than $\kappa^e_{zz}$. While the in-plane $\kappa_l$ is comparable to $\kappa_e$. We fit the experimental data by using $\kappa$ = 183.103/T + 0.412671. It seems that Umklapp process which usually has $\kappa$ $\sim$ 1/T shows up and thermal conductivity gets softening with increasing temperature.

All the data obtained above allow us to calculate ZT as function of temperature and hole doping along two directions, which is shown in Fig.~\ref{Fig6}(a-d). Optimum ZT value increases with increasing temperature, so does the corresponding optimum doping level. In-plane is better than cross-plane for thermoelectric conversion, with its ZT up to 0.30 and saturated around 700 K, as shown in Fig.\ref{Fig6}(a,b). We may further reduce $\kappa$ by random stacking according to Kim \cite{kappa2} and Chiritescu et. al. \cite{science315.351}, but its effect on electrical transport needs to be checked if it may compromise the gain of ZT by reducing thermal conductivity.

\section{CONCLUSION}
By combining ab. initio. band structure calculation with semi-classical Boltzmann transport theory, we theoretically studied the doping and temperature dependence of thermoelectric transport properties of 2H-MoS$_2$. Anisotropic electronic scattering time has to be considered to account for difference between in-plane and cross-plane electrical conductivity $\sigma$, which also gives rise to anisotropy of electronic thermal conductivity $\kappa_e$. In-plane $\kappa^{xx}_e$ is comparable to lattice $\kappa^{xx}_l$ while cross-plane lattice $\kappa^{zz}_l$ dominates over lattice $\kappa^{zz}_e$. In contrast to the anisotropy of $\sigma$ and $\kappa_e$, thermopower, which is attainable more than 200 $\mu$V/K over a wider range of doping and temperature, shows a vanishing anisotropy for doping over 10$^{17}$ cm$^{-3}$. The maximum ZT can reach as high as 0.3 at around 700 K with carrier density of 10$^{20}$ cm$^{-3}$, and may go higher if restacking process is used to further reduce the thermal conductivity. A preference for the in-plane thermoelectric transport by doping is demonstrated.

\acknowledgments
This work was supported by the NSFC under Grant No. 11004201, 50831006 and the National Basic Research Program (No. 2012CB933103). T.Y. acknowledges the IMR SYNL-Young Merit Scholars and T.S. K$\hat e$ research grant for support.

\end{document}